\documentclass[9pt,twocolumn,twoside]{osajnl}

\journal{ol} 

\usepackage[colorinlistoftodos]{todonotes}
\usepackage{graphicx}
\definecolor{red}{rgb}{1, 0, 0}

\setboolean{shortarticle}{true} 

\ifthenelse{\boolean{shortarticle}}{\colorlet{color2}{color2b}}{\colorlet{color2}{color2}} 

\title{Non-destructive shadowgraph imaging of ultracold atoms}

\author[1,*]{P. B. Wigley}
\author[1]{P. J. Everitt}
\author[1]{K. S. Hardman}
\author[2]{M. R. Hush}
\author[1,3]{C. H. Wei}
\author[1]{M. A. Sooriyabandara}
\author[1]{Manju P.}
\author[1]{J. D. Close}
\author[1]{N. P. Robins}
\author[1]{C. C. N. Kuhn}

\affil[1]{Quantum Sensors and Atomlaser Group, Department of Quantum Science, Research School of Physics and Engineering, The Australian National University, Acton, 2601, Australia}
\affil[2]{School of Engineering and Information Technology,
University of New South Wales at the Australian Defence Force Academy, Canberra, 2600, Australia}
\affil[3]{Department of Instrument Science and Technology, College of Mechatronic Engineering and Automation, National University of Defense Technology, Changsha 410073, China}
\affil[*]{Corresponding author: paul.wigley@anu.edu.au}

\dates{Compiled \today}

\ociscodes{(020.1475) Bose-Einstein condensates; (070.0070) Fourier optics and signal processing; (070.2580) Paraxial wave optics}

\doi{}

\setboolean{displaycopyright}{false}

\begin{document}
\begin{abstract}
An imaging system is presented that is capable of far-detuned non-destructive imaging of a Bose-Einstein condensate with the signal proportional to the second spatial derivative of the density. Whilst demonstrated with application to $^{85}$Rb, the technique generalizes to other atomic species and is shown to be capable of a signal to noise of ${\sim}25$ at $1$GHz detuning with $100$ in-trap images showing no observable heating or atom loss. The technique is also applied to the observation of individual trajectories of stochastic dynamics inaccessible to single shot imaging. Coupled with a fast optical phase lock loop, the system is capable of dynamically switching to resonant absorption imaging during the experiment.
\end{abstract}
\maketitle
\thispagestyle{fancy}

\ifthenelse{\boolean{shortarticle}}{\ifthenelse{\boolean{singlecolumn}}{\abscontentformatted}{\abscontent}}{}

Experiments using ultra-cold gases play an important role in the study of quantum physics, with Bose-Einstein condensates (BECs) forming the archetypal quantum system.
Typically the time spent generating this resource far outweighs that spent on the corresponding experiment. Reduction of this duty cycle is key to improving many experiments including the sensitivity of cold atom based sensors \cite{robins_atom_2013}. Continuous imaging provides one such way to effectively improve duty cycle, allowing quantum resources to be probed multiple times in a single experimental run. Additionally, a continuous imaging system allows experiments to probe regimes inaccessible to traditional single shot imaging, including stochastic process such as dynamic instability of solitons confined to an optical waveguide \cite{carr_spontaneous_2004,leung_nonlinear_2002}.

Despite the comparative difficulty of producing a BEC, many experiments still use single shot absorption imaging to acquire experimental data. Absorption imaging is robust and gives excellent signal-to-noise ratios (SNR) for the atomic densities typically encountered in experiments, but is totally destructive. Conversely, dispersion based imaging can provide quasi-continuous minimally destructive imaging, albeit with lower SNR. Dispersive imaging techniques have been used extensively to observe thermal clouds and BEC alike \cite{bradley_analysis_1997,bradley_bose-einstein_1997,gajdacz_non-destructive_2013}. Dark-ground imaging \cite{andrews_direct_1996,pappa_ultra-sensitive_2011,reinhard_dark-ground_2014} is a common dispersive technique where the non-interacting light is blocked in the Fourier plane. A direct improvement instead involves inserting a phase plate at the Fourier plane in order to increase the SNR \cite{andrews_propagation_1997,meppelink_thermodynamics_2010}. Both techniques involve the precise alignment of optics and preclude conventional absorption imaging from being implemented on the same optical path. Other dispersive methods have been developed to exploit atomic birefringence to measure a rotation of polarization \cite{higbie_direct_2005,gajdacz_non-destructive_2013,franke-arnold_magneto-optical_2001} but require magnetic fields to operate. Phase contrast imaging is widely used in xray tomography \cite{langer_quantitative_2008} with the signal described in both the near-field and far-field regime using the transport of intensity equation (TIE) and contrast transfer function (CTF) respectively. In the far-field, the CTF approach is used to solve the inverse problem and requires multiple images along the beam propagation direction. Despite this requirement, the technique has been applied to cold atoms \cite{turner_diffraction-contrast_2005}. The near-field signal can be reimaged using a single lens placed at $2f$ or a telescope setup, with the TIE approach relaxing the requirement for multiple images. This approach has previously been used to image a magneto-optical trap \cite{turner_off-resonant_2004} with low SNR and high destruction. We extend the approach to image a BEC in-trap in the far-detuned regime, with no observable heating or atom loss over $100$ images. The signal is shown to be dependent on the second spatial derivative of the refractive index, as in the shadowgraph technique first observed by Hooke \cite{hooke_new_1665,settles_schlieren_2001}. 

The imaging combines three technologies: (i) a low cost, high quantum efficiency, high speed CMOS camera, (ii) an imaging system simultaneously capable of far detuned dispersive imaging and near resonant absorption imaging, and (iii) a recently developed, low cost, high performance OPLL \cite{wei_compact_2016} that allows for dynamic tuning of the imaging laser over GHz, allowing both phase contrast and absorption imaging in a single run. The performance of the system is characterized and demonstrated using a dual species $^{87}$Rb/$^{85}$Rb BEC apparatus \cite{kuhn_bose-condensed_2014}. $^{85}$Rb provides a powerful playground for such an imaging system as it experiences a Feshbach resonance at 155G. This resonance allows for direct manipulation of the nonlinearity of the Gross-Pitaevskii equation and provides access to experiments such as solitonic propagation and collisions \cite{nguyen_collisions_2014,mcdonald_bright_2014}, and the bosenova \cite{saito_intermittent_2001,saito_power_2001,donley_dynamics_2001,altin_collapse_2011}. Typical $^{85}$Rb condensates contain a significantly lower atom number than other species due to unfavorable inelastic scattering cross-sections resulting in a lower imaging signal. Despite this, the imaging system is shown to be capable of producing high SNR images of the sample. A number of examples demonstrate the system, including dual atomic species imaging, continuous imaging of small $^{85}$Rb solitons (${\sim}10^3$ atoms), large $^{87}$Rb condensates (${\sim}10^6$ atoms), continuous acquisition of data over the entire experimental duty cycle (images spaced by $100$ms), and dynamical data acquisition where a sequence of phase contrast images are followed by a high SNR near-resonant absorption image. The low cost components, coupled with the high performance and utility of this imaging setup make it a powerful tool for ultracold atom experiments.

\begin{figure}[t]
\centering{}
 \includegraphics[width=8.5cm]{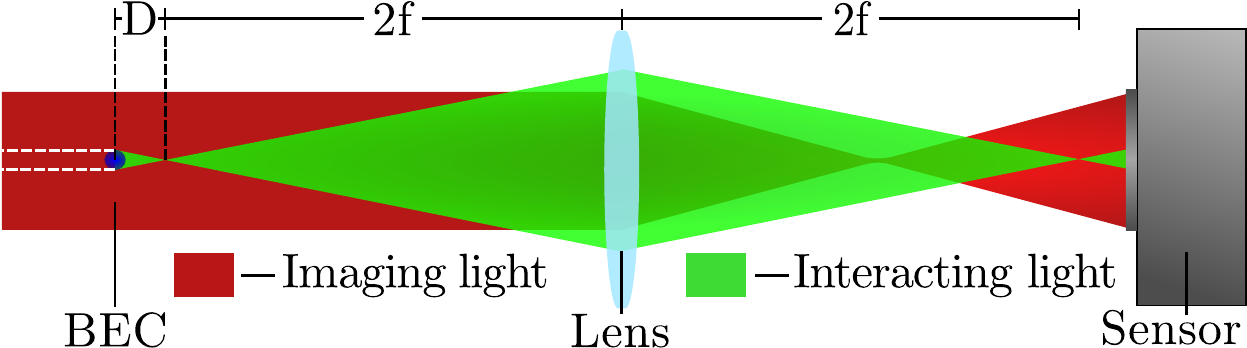}
 \caption{Simplified schematic of the imaging system.The probe light interacts with the BEC, acquiring a phase shift and producing a change in the spatial mode of the light which is subsequently detected on a high frame-rate complementary metal-oxide semiconductor (CMOS) camera.}
 \label{scheme}
\end{figure}

Imaging relies on the interaction between light and the sample of atoms, dictated by the refractive index \cite{ketterle_making_1999},
\begin{align}
n\left(\textbf{x},\,z\right)=&1 + \frac{\sigma_0 \lambda\,\rho\left(\textbf{x},\,z\right)}{ 4\pi}\left(\frac{i}{1+\delta^2} -\frac{\delta}{1+\delta^2}\right),
\label{refractiveindex}
\end{align}
where $\sigma_0=3\lambda^{2}/2\pi$ is the cross-section, $\lambda$ is the wavelength, $\rho$ is the atomic density, $\delta = \Delta/(\Gamma/2)$ is the detuning ($\Delta$) in half linewidths ($\Gamma/2$) and $\textbf{x}=\{x,\,y\}$ is the spatial coordinates of the plane perpendicular to the direction of propagation.
Key to this expression are the distinct real and imaginary components, manifesting physically as dispersive and destructive signals respectively. 
Any probe beam passing through the atoms will be attenuated and phase-shifted with the resultant field written as 
\begin{eqnarray}
E(\textbf{r})=E_{0}\exp{\left(\frac{2\pi}{\lambda}\int \left(n\left(\textbf{r},\,z\right) -1\right)dz\right)} =E_{0}t(\textbf{r})\textrm{e}^{i \phi\left(\textbf{r}\right)}.
\end{eqnarray}
At resonance the real part of the refractive index is zero, resulting in no phase shift ($\phi=0$) and the imaginary part of the refractive index solely contributing to the signal in the form of absorption. As the probe beam is detuned, the real part of the refractive index becomes non-zero and generates a phase shift in the light with the two signals described by
\begin{align}
t(\textbf{r})&=\textrm{exp}\left(-\frac{\sigma_0\tilde{\rho}(\textbf{x})}{2}\frac{1}{1+\delta^2}\right)\\
\phi(\textbf{r})&=\frac{\sigma_0\tilde{\rho}(\textbf{x})}{2}\frac{\delta}{1+\delta^2},
\end{align}
where $t(\textbf{r})$ and $\phi(\textbf{r})$ corresponds to the absorption and phase shift respectively, with both components of the refractive index taken relative to vacuum and $\tilde{\rho}(\textbf{x}) = \int n\cdot dz$ the column density. 
The pattern produced by the interaction with the sample is one of Fresnel diffraction with the transmittance altered according to the Fresnel propagator \cite{goodman_introduction_2005},
\begin{eqnarray}
P_{D}\left(\textbf{x}\right) = \frac{1}{i\lambda D}\exp{\left(i\frac{\pi}{\lambda D}\left|\textbf{x}\right|^2\right)}
\label{fresnelPropagator}
\end{eqnarray}
where $D$ denotes the distance of propagation along $z$. The resultant intensity pattern at $D$ is given by the modulus square of the transmittance propagated with the Fresnel propagator,
\begin{eqnarray}
I_D\left(\textbf{x}\right) = \left|E\left(\textbf{x}\right)\ast P_D\left(\textbf{x}\right)\right|^2.
\label{convolutionPropagator}
\end{eqnarray}
This propagation becomes simpler when considered in the Fourier domain with the diffraction pattern being given by
\begin{eqnarray}
\tilde{I}_{D}(\textbf{f}) = \int E\left(\textbf{x}-\frac{\lambda D\textbf{f}}{2}\right)E^{\ast}\left(\textbf{x}-\frac{\lambda D\textbf{f}}{2}\right)\exp{\left(-i 2\pi\textbf{x}\cdot\textbf{f}\right)}d\textbf{x}.
\label{intensityFourier}
\end{eqnarray}
\begin{figure}[t!]
 \centering{}
  \includegraphics[width=\columnwidth]{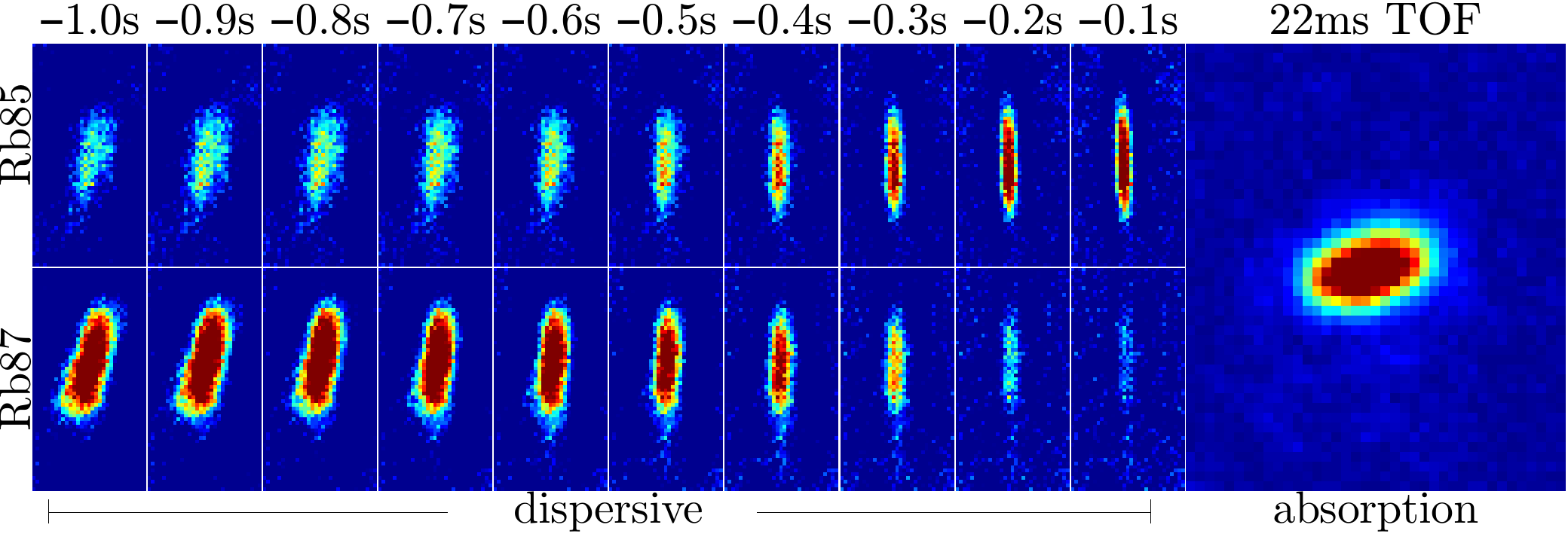}
  \caption{Sympathetic evaporation leading to BEC in $^{85}$Rb. Top row shows $10$ in-trap images of $^{85}$Rb, $100$ms apart with the final image taken $100$ms before condensation. The lower instead images $^{87}$Rb. Sympathetic evaporation cools the sample, with the $^{87}$Rb content reducing to zero as the $^{85}$Rb atoms become cooler and denser. The rightmost picture shows a $22$ms time of flight absorption image confirming the phase transition and an accurate count of atoms number, $3\times10^4$.}
 \label{Evaporation}
\end{figure}
Given a diffraction pattern, the inverse problem must be solved to relate this to the density of the sample. In the small $D$ regime, and in the limit of small, flat absorption ($t(\textbf{r})\approx1$), the TIE approach can be used an the intensity pattern related to the phase by \cite{langer_quantitative_2008}
\begin{eqnarray}
I_{D}\left(\textbf{x}\right) = I_0\left[1-\frac{\lambda D}{2\pi}\boldsymbol{\nabla}^2\phi\left(\textbf{x}\right)\right],
\end{eqnarray}
where $I_0=|E_0|^2$. This relationship is one of a number of solutions to solving the inverse problem including partial differential equation methods \cite{allen_phase_2001,gureyev_hard_1999,gureyev_rapid_1997,gureyev_phase_1996,teague_deterministic_1983,langer_quantitative_2008} and Fourier methods \cite{paganin_noninterferometric_1998}. The condensate effectively lenses the incident light with the optimal signal occurring at the focal point of this lens ($D$ in figure \ref{scheme}). Since the condensate is dense, the effective focal length is short compared to the imaging plane allowing the sensor to be placed in-focus for absorption, enabling fast switching between imaging methods.

The experimental apparatus used to produce the $^{87/85}$Rb BECs has been described in detail in \cite{kuhn_bose-condensed_2014}. Briefly, a magneto-optical trap (MOT) is loaded with both atomic species. $25$ms of polarization gradient cooling is applied, resulting in a ${\sim}15\mu$K sample and both isotopes pumped to their respective magnetic ground states. Using a hybrid magnetic and optical trap, the $^{85}$Rb atoms are then sympathetically cooled with $^{87}$Rb atoms. The remaining cloud, cooled to around $1\mu$K, is transferred to an optical crossed dipole trap where it is cooled further by reducing the dipole beam intensity, driving sympathetic evaporative cooling until the BEC phase transition is reached, as indicated in Figure \ref{Evaporation}. The cloud may be loaded into an optical waveguide beam by extinguishing one of the dipole beams. 

The $^{85}$Rb atomic interactions are manipulated using a magnetic bias field through the Feshbach resonance. The field is jumped through the resonance at $155$G to $165.74$G \cite{roberts_improved_2001}, minimizing inelastic collisional losses. During the last $0.5$s of evaporation, the bias field is tuned such that the scattering length of the $^{85}$Rb atoms is $254a_0$, increasing the physical size in order to optimize atom number. Pure $^{85}$Rb and $^{87}$Rb can be produced by changing the ratio of the two species in the MOT loading. 

\begin{figure}[t!]
\centering{}
 \includegraphics[width=\columnwidth]{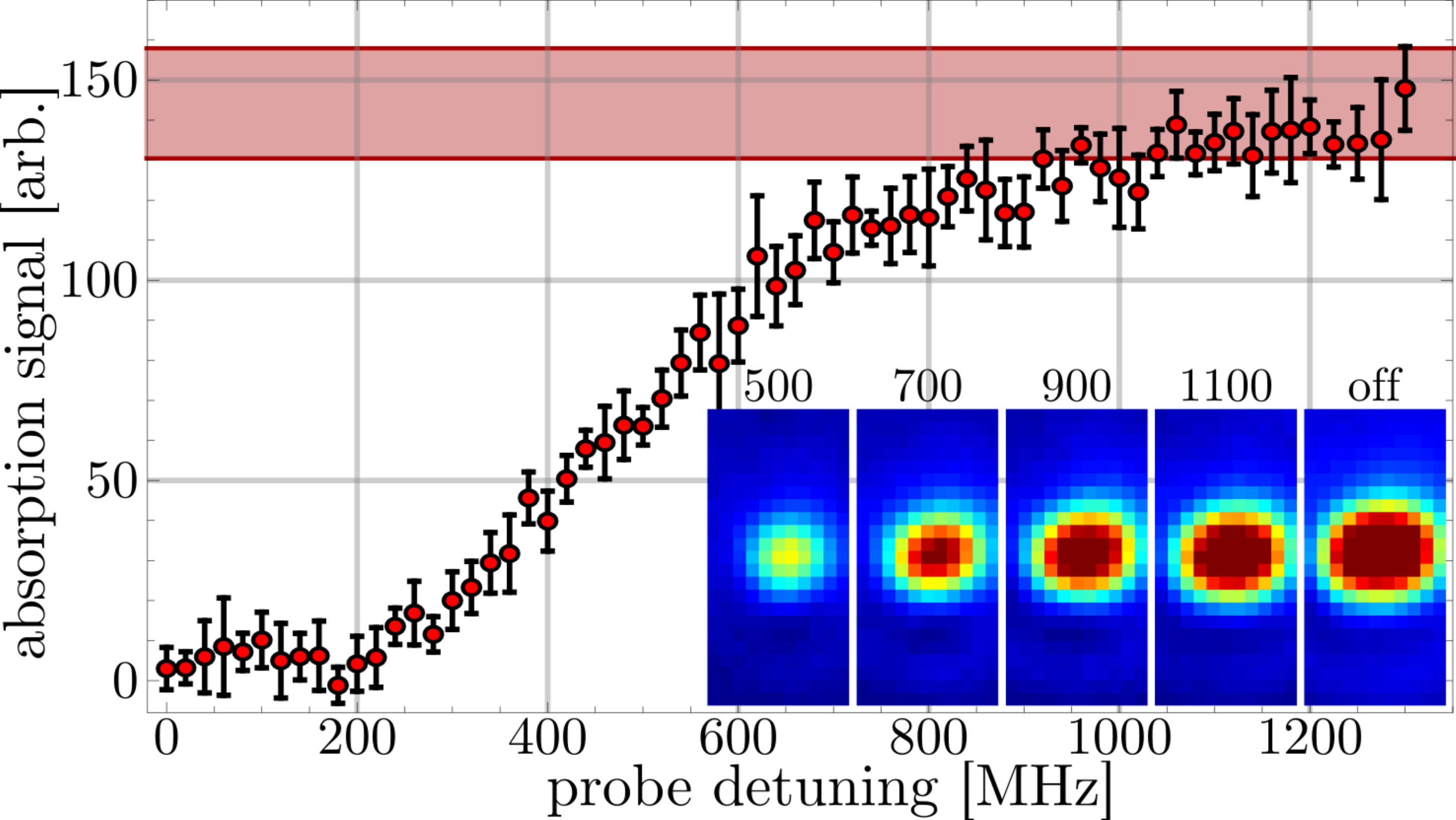}
 \caption{Integrated signal from the $22$ms TOF absorption system following $100$ dispersive in-trap images spaced $2$ms apart. Maximum destruction occurs at resonance and lessens as the in-trap probe is detuned further. The shaded area indicates one standard deviation either side of the mean integrated signal when the probe is off. The signal is seen to approach this region as the detuning approaches $1$GHz. Insets show the signal in the TOF absorption image at various detunings from resonance, which at large detuning show negligible difference.}
 \label{dropSignal}
 \end{figure}

To achieve the required detuning of the imaging probe beam, a dedicated external cavity diode laser with optical phase locking loop (OPLL) is used. 
A master laser locked to an atomic transition using saturated absorption spectroscopy is beat with the probe laser on a $12$GHz fast photo detector. 
The signal is amplified and divided down such that it can be mixed with a radio frequency source. The beat lock and feedback is generated using the method described in \cite{wei_compact_2016} with the feedback applied on the probe laser using piezo control. The phase lock loop provides lock with narrow linewidth below $1$Hz and a capture range of $10$MHz enabling swift changes of detuning and dynamic switching between dispersive and absorption imaging. 

The probe beam is delivered to the science table by a polarization maintaining fiber where it is combined with the vertical MOT beam, sharing the same optical path through the science cell. This orientation allows for imaging perpendicular to the optical waveguide beam, allowing for non-destructive probing of experiments performed in the waveguide. Two liquid crystal wave-plates enable dynamic switching of polarization of this optical line, allowing different polarizations of light for the MOT and imaging beams. A dual lens ($f=10$cm) telescope allows the near-field signal to be reimaged far from the vacuum system. Finally a $4$ times magnification is achieved using an objective lens with the subsequent signal detected on a CMOS camera (Point Grey GS3-U3-41C6NIR-C). The camera has high quantum efficiency ($45\%$ at $780$nm) and is capable of kHz frame rates. The setup enables fast, dynamic imaging that is minimally destructive despite a small and fragile sample and is seen to be robust to extra optics sharing the same path.  

\begin{figure}[t]
 \centering{}
 \includegraphics[width=8.5cm]{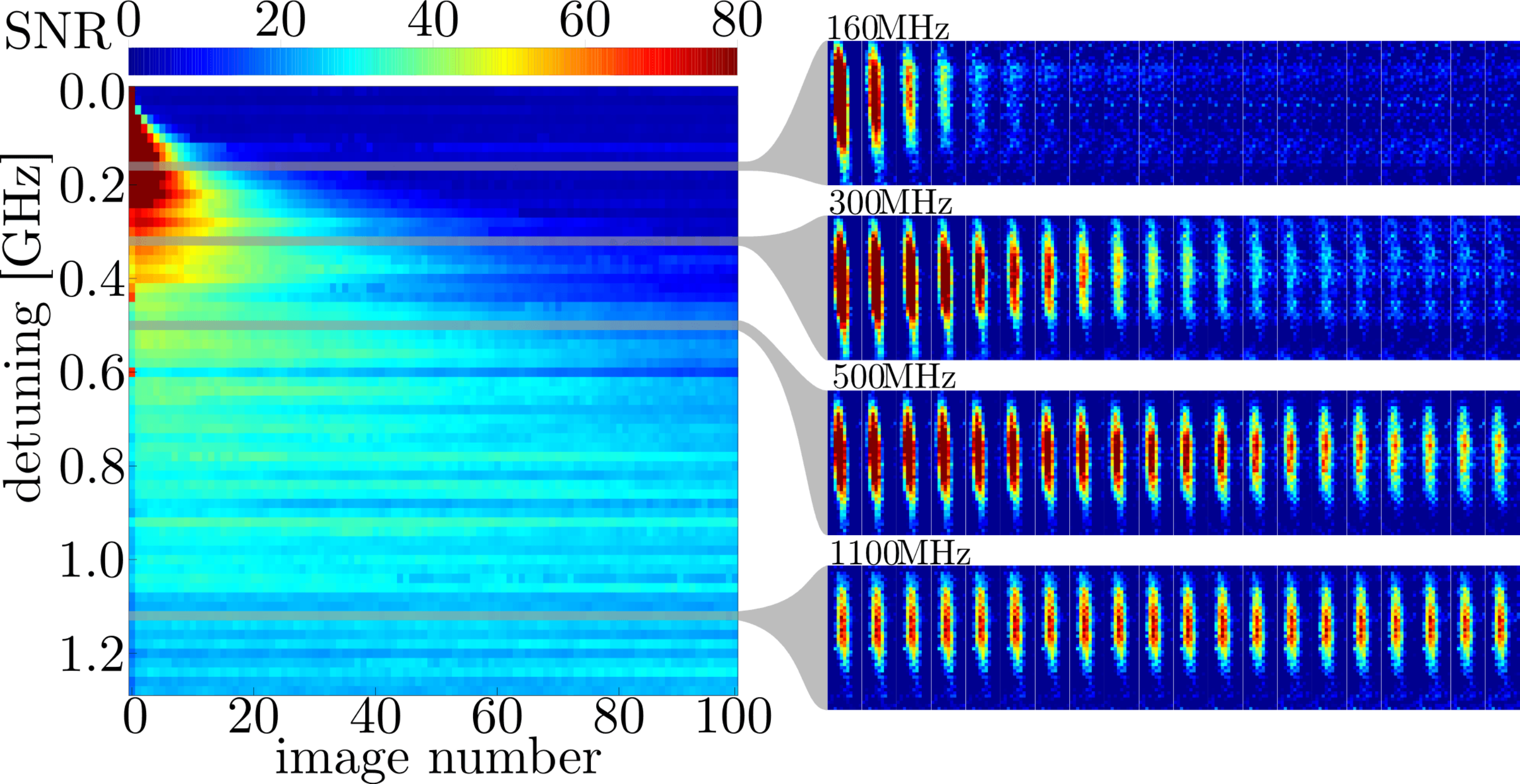}
 \caption{Left: The SNR for each of the $100$ in-trap images for a range of detunings. A Fourier DCT filter is applied and the images averaged over $10$ sets. Two regions are selected, one around the peak signal and another in the background. Each region is integrated along one axis before a cumulative second integration is taken. The signal is given by the final value in the second integration. The noise is determined from the variance in the second integration of the background. The SNR is largest at resonance, but decays quickly as the condensate is destroyed. As the detuning is increased the SNR decreases as does the destruction with constant SNR (${\sim}20$) seen for large detunings. Right: $20$ image subset of the $100$ in-trap images over a period of $200$ms, shown for various detunings.}
 \label{combinedSNRImagesSmaller}
\end{figure}
The imaging system was tested over a range of detunings, timescales and destruction. A $^{85}$Rb condensate was probed through $100$ in-trap images over a GHz range of detuning, with each image taken $2$ms apart. The condensate was then dropped and imaged after a $22$ms time-of-flight by a separate horizontal absorption system. Equally, the OPLL could have been used to move the probe detuning on resonance for the final image. 

Figure \ref{dropSignal} outlines the change in the condensate as the detuning of the dispersive imaging changes. Close to resonance the condensate is completely destroyed. As the detuning increases this destruction decreases. The shaded band at the top of the figure indicates one standard deviation from the mean variation run-to-run with no probe applied. The absorption signal is seen to approach this regime for large detunings with no change in atom number or cloud widths observed after expansion. Even far detuned the in-trap signal remains high, as evidenced in the right of Figure \ref{combinedSNRImagesSmaller}, where $20$ image subsets are shown for a range of detunings. A Fourier discrete cosine transform filter has been applied to each image and each has been averaged over series of $10$ runs for each detuning. In application to stochastic processes, such averaging is unavailable, and the signal drops accordingly. At ${\sim}1$GHz the signal has negligible decay and the corresponding dropped cloud closely resembles the cloud with no in-trap probe beam. 
\begin{figure}[t!]
\centering{}
 \includegraphics[width=\columnwidth]{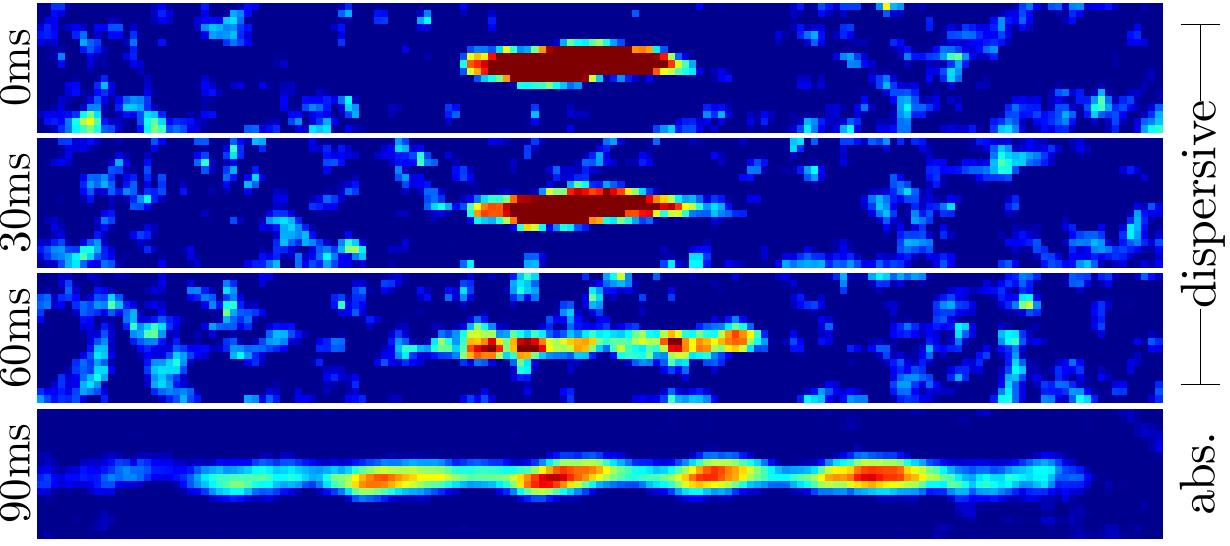}
 \caption{Dynamic manipulation of in-trap probe detuning. The $4$ images are a subset of $9$ in-trap images with the final taken on resonance. The system allows direct observation of a soliton confined to an optical waveguide decaying into a soliton train, a stochastic event with the breakup time and number of components varying run-to-run. Continuous imaging enables observation of individual trajectories of stoachastic processes.}
 \label{MIGrid}
 \end{figure}
Integrating an area of the in-trap image where no signal is present allows a characterization of the noise. The area of integration is the same for both the signal and the noise. The resultant SNR is shown in the left of Figure \ref{combinedSNRImagesSmaller}. The peak signal is close to resonance, however it decays exponentially in time as the condensate is destroyed. As the detuning is increase the signal decreases, but so too does the loss rate due to absorption. Far from resonance, at $1$GHz, negligible decay occurs and the SNR is ${\sim}25$ and corresponds to near optimal operation in terms of both destruction and signal. 

A final demonstration is shown in Figure \ref{MIGrid} where a soliton confined to an optical waveguide undergoes stochastic breakup. The onset time and number of components in the resultant soliton train varies stochastically preventing single-shot imaging systems from capturing the dynamics. Continuous probing allows these dynamics to be observed analyzed analysis.

A fast, dynamically tunable system capable of imaging small samples with negligible destruction has been presented. The technique generalizes to other atomic species and is shown to be capable of imaging a small sample of only $10^4$ atoms up to $100$ times with high signal to noise (${\sim}25$) and negligible decrease in atom number and no observable heating. With the ability to dynamically change the detuning of the probe beam, the dispersive imaging can be followed by a high signal to noise absorption image for accurate analysis. The performance was characterized and a number of applications presented, demonstrating the benefit of such a system to any cold atom experiment.

The authors gratefully acknowledge the support of the Australian Research Council (ARC) Discovery program. M.R.H acknowledges support of ARC Discovery Project DP140101779. 







\bibliography{nondestructive}


\end{document}